\documentstyle[12pt]{article}

\begin{document}
\begin{center}

{\large \bf Temperature induced alignment in hot rotating nuclei}\\
\vspace{0.5 true in}

B. K. Agrawal$^1$  and A. Ansari\\

{\small Institute of Physics, Bhubaneswar-751005, India}\\
\end{center}
\footnotetext[1]{e-mail: bijay@iopb.ernet.in}
\begin{center}
ABSTRACT
\end{center}

{\small Employing a cranking Hamiltonian, in the static path approximation
to the partition function, we have studied the behaviour of the
moment of inertia of $^{64}Zn$ at a few values of temperatures
as a function  of the rotation frequency. At a low temperature
T = 0.5 MeV and J=16 rotational alignment of $0g_{9/2}$ orbitals
takes place. But at a high temperature T = 2.0 MeV about 40 - 45$\%$
of all the angular momenta are generated by the alignment of the
same orbitals. This should imply a decrease in collectivity for all
the spins at high temperatures.}
\vspace{0.5 true in}


\newpage
Recently it has been realised that for the description of hot nuclei one
must go beyond a mean field approach as thermal fluctuations in the
shape degrees of freedom become very important at high temperatures[1].
The so called static path approximation(SPA)[2-5] to the path integral
representation of partition function[6] provides such a scheme with some
limitations at low temperatures $T<$ 1.0 MeV[7,8]. Recently we[5] have
made application of the SPA approach to a somewhat realistic situation for
some zinc isotopes using a quadrupole interaction Hamiltonian in a reasonable
model space. Pairing interaction has not yet been included and calculations
are performed for $T\ge$ 0.5 MeV. There are no experimental data available
for these nuclei. However, as discussed in Ref. [5], the calculated numbers
for level density, $\rho$ and level density parameter $a$ etc. look
reasonable with some new features. For instance, we find that $ln\rho$
increases with the increase of T, somewhat faster than expected, even
within the small range of T= 0 - 2.5 MeV.

To bring in spin dependence to the level density we have employed the
same Hamiltonian with a $J_x$-cranking term added to it. That is we have
\begin{equation}
\hat H^\omega=\hat H-\omega \hat J_x
\end{equation}
\noindent where $J_x$ is the x-component of the angular momentum operator
with $\omega$ as the cranking (rotational) frequency. The path integral
representation of the partition function, Z, in SPA is given by
\begin{equation}
Z(\mu_p,\mu_n,\omega,T)=4\pi^2({\alpha\over 2\pi T})^{5/2}
\int_o^\infty \beta^4 d\beta\int_o^{\pi/3}d\gamma sin3\gamma
e^{-{\alpha\beta^2\over 2T}}Tr e^{-(\hat H^{\prime\omega}-\mu_p\hat N_p-
\mu_n\hat N_n)/T}
\end{equation}
\noindent $\hat H^{\prime\omega}=\sum_i h^{\prime\omega}(i)$, where
$h^{\prime\omega}$ is a Nilsson type deformed mean field Hamiltonian in the
rotating
frame.
\begin{equation}
\hat h^{\prime\omega}=h_o-\hbar \omega_o\beta {r^2\over b^2}[cos\gamma
Y_{2o}+{1\over \sqrt 2}sin\gamma(Y_{22}+Y_{2-2})]-\omega \hat j_x
\end{equation}
\noindent with $h_o$ representing the spherical basis sp energies defined with
respect to an appropriate inert core.  The value of $\hbar \omega_o=41A^{-1/3}$
MeV and $\alpha=(\hbar \omega_o)^2/\chi b^4$ with $\chi b^4$ = $70A^{-1.4}$
MeV. The integration variables $\beta$ and $\gamma$ in (2) are
the well known quadrupole deformation parameters as given by Baranger and
Kumar[9] (for the rare earth nuclei).  Here $\chi$ is the quadrupole
interaction
strength and $b$ the usual oscillator length parameter. The chemical potentials
$\mu_p$ and $\mu_n$ are adjusted to reproduce the proton and neutron numbers,
$N_p$ and $N_n$, respectively. The particle numbers
\begin{equation}
N_{p,n}=T{\partial lnZ\over \partial \mu_{p,n}}
\end{equation}
\noindent Similarly the angular momentum constraint is met by adjusting
$\omega$ such that
\begin{equation}
\sqrt{J(J+1)}=<J_x>=T{\partial lnZ\over \partial \omega}
\end{equation}
\noindent Finally a moment of inertia can be defined as
\begin{equation}
\Im_J={\sqrt{J(J+1)}\over\omega_J}
\end{equation}
\noindent In the above $\omega_J$ represents the final vlaue of $\omega$
which satisfies the constraint (5).

With the help of the partition function (2) various quantities like free energy
and level density $\rho$ etc. can be computed. But here we are mainly
interested to study the variation of $\Im_J$
as a function of $\omega_J$ at a few temperatures for a nucleus $^{64}Zn$.
We may remind here that  such a study can be done in a mean field, like cranked
Hartree-Fock-Bogoliubov, theory. But here the advantage is that there is
an integration over the $\beta$ and $\gamma$ parameters correcting
for the shape fluctuations at a finite temperature. As will be seen below
the
variation of $\Im_J$ in the T - J (or $\omega_J$) plane reveals some
interesting structural information.

The computational details are given in Ref. [5]. In order to have
a good number of active particles we have chosen Z = N = 20, i.e.,
$^{40}Ca$ as an inert core. Thus for $^{64}Zn$ there are 10  protons and 14
neutrons each in 30 orbitals in the model basis space, extended up to
$0g_{9/2}$. The spherical basis sp energies are -14.4, -10.2, -8.8, -8.3
and -4.4 (all in MeV) for the orbitals $0f_{7/2}$, $1p_{3/2}$,
$0f_{5/2}$, $1p_{1/2}$ and $0g_{9/2}$, respectively. These values are
precisely those given by Lauritzen and Bertsch[3]. As a first step to the
trace calculation in Eq. (2) the deformed cranked sp Hamiltonian of
Eq. (3) is diagonalised in the full basis space at $\beta$-$\gamma$
mesh points. Finally $\beta$, $\gamma$-integrations are performed using
Gaussian quadrature methods.

According to Eq. (6) the moment of inertia is a ratio of the
angular momentum to the rotational  frequency (J$\sim \omega\Im$),
rather a classical relation. But since $<J_x>$ is computed in a fully
microscopic manner, the moment of inertia is actually not solely a parameter
describing collective rotation of the nucleus. It is well known from
the yrast state high-spin studies[10] that the total angular momentum
can be generated from collective rotations as well as alignment of
a few high-j sp orbitals. The cranking procedure[11,12] is known
to incorporate both of these mechanisms. In Fig. 1 we have shown the
variation of moment of inertia as a function of $\omega$ at a few
temperatures T = 0.5, 1.0, 1.5 and 2.0 MeV. At every indicated
point of the curves the angular momentum constraint (5) is satisfied.
At a low temperature T = 0.5 MeV, the plot shows a rather sudden rise
in the moment of inertia at $\omega$ $\sim$ 1.0 MeV and J$\approx$
16. This implies the generation of a substantial part of the angular momenta
$J\ge$ 16 by rotation aligments of low-m and high-j sp orbitals which
happens to be $0g_{9/2}$ in the present case. Denoting the contribution
of a few particles (protons as well as neutrons) in $0g_{9/2}$
as aligned angular momentum $j_a$, and the remaining fraction as
collective $J_{coll}$, one can write
\begin{equation}
<J_x>=J_{coll}+j_a
\end{equation}
\noindent With the increase of temperature $\Im_J$ increases even at the
lowest spin J = 2. Finally, for example, at T = 2.0 MeV the moment of
inertia becomes almost a constant and independent of angular momenta.
Another interesting feature visible is that $\Im_J$ at J = 16 is almost
temperature independent. That is at J = 16 the contribution $J_{coll}$
and $j_a$ are almost the same at all temperatures ( a sort of saturation
of the alignment at J = 16 and T$\sim$0 itself). In order to gain further
insight in these features we have plotted in Fig. 2 the relative contribution
$j_a/<J_x>$ coming from $0g_{9/2}$ orbitals and the number of particles
(protons
+ neutrons) as a function of J at two temperatures T = 0.5 and  2.0 MeV.
At T = 0.5 MeV and J = 2 there are hardly any particles occupying
$0g_{9/2}$ orbitals and the value of $j_a$ is negligible. At T = 2.0 MeV
and J = 2 itself there are on the average about 1.5  particles in  $0g_{9/2}$
and these contribute more than 40$\%$ to the total spin. This we may
term as caused by a temperature induced alignment. Furthermore, at T = 0.5
MeV the occupancy of $0g_{9/2}$ orbitals increases with the increase
of J correspodingly increasing the value of the contribution $j_a$
to the total J. By J = 16 - 18, this alignment essentially saturates( it
may be noted that 2 protons + 2 neutrons in $0g_{9/2}$ orbitals could
contribute maximum 16 units). On the otherhand at T = 2.0 MeV the
contribution $j_a/<J_x>$, which is plotted as percentage, has
saturated for all angular momenta.

To conclude, these results  should have important bearings at high
temperature spectroscopic studies. Depending upon the availability
of high-j sp intruder orbitals near the Fermi surface, the angular
momenta will be generated to a great extent by a temperature induced
alignment at high T and for relatively low spins. It implies
a decrease in collectivity with the increase of temperature even
at low spins. This should be noted keeping in mind the shape
fluctuation integrations in the $\beta-\gamma$ space in Eq. (2).
The experimental confirmation of such a behaviour would be highly
interesting, though  for the residual compound nucleus in a low spin
and highly excited state it may be hard to make spectroscopic observations.

\begin{figure}[pt]
\caption {Variation of moment of inertia with cranking frequency
$\omega_J$ at T = 0.5 - 2.0 MeV}
\caption {Variation of N and $j_a$ contributions from $0g_{9/2}$
orbitals as functions of J and T. As indicated the scale for $j_a$ is
given on the left hand side whereas for the number N is on the right hand side.
\end{figure}

\begin{thebibliography}{99}

\bibitem{} Y. Alhassid, B. Bush and S. Levit, Phys. Rev. Lett. 61, 1926
(1988)
\bibitem{} B. Lauritzen, P. Arve and G. Bertsch, Phys. Rev. Lett. 61, 2835
(1988)
\bibitem{} B. Lauritzen and G. Bertsch, Phys. Rev. C39, 2412 (1989)

\bibitem{} Y. Alhassid and B. Bush, Nucl. Phys. A549, 43 (1992)
\bibitem {} B. K. Agrawal and A. Ansari, Phys. Rev. C46, 2319 (1992)

\bibitem{} J. W. Negele and H. Orland, {\it Quantum Many Particle Systems}
(Addison-Wesley, Reading, MA, 1988)
\bibitem{} G. Puddu and P. F. Bortignon and R. A. Broglia, Phys. Rev. C42,
R1830 (1990)
\bibitem{} R. Rossignoli, P. Ring and N. Dinh Dang, Phys. Lett. B297,
9 (1992)
\bibitem{} N. Baranger and K. Kumar, Nucl. Phys. A110, 490, 529 (1968)
\bibitem{} D. Schwalm, Nucl. Phys. A396, 339c (1983)
\bibitem{} A. Ansari and S. C. K. Nair, Phys. Rev. C32, 637 (1985)
\bibitem{} M. Ploszajczak and A. Faessler, Nucl. Phys. A379, 77 (1982)
\end{thebibliography}
\end{document}